\documentclass[12pt]{article}
\usepackage{graphicx}
\usepackage{amsmath}
\usepackage{amsfonts}
\usepackage{amssymb}

\begin{document}

\title{\bf Markov-chain approach to a process with long-time memory}

\author{Guglielmo Lacorata\\Dipartimento di Fisica, Universit\'{a} di Roma 
``La Sapienza'' Italy, and\\
  Dipartimento di Fisica, Universit\'{a} de L'Aquila, Italy\\
\and Rub\'{e}n A. Pasmanter\\Royal Dutch Meteorological Institute, 
Netherlands\\
 \and Angelo Vulpiani\\Dipartimento di Fisica, INFM and SMC,\\ 
Universit\'{a} di Roma ``La Sapienza'', Italy\\
}

\maketitle

\thispagestyle{empty}
\newpage
\setcounter{page}{1} 
\pagestyle{plain}

\begin{abstract}
\noindent 
We show that long-term memory effects, present in the chaotic
dispersion process generated by a meandering jet model, can be 
nonetheless taken into account 
by a first order Markov process, provided that the 
states of the phase space partition, chosen to describe the
system, be appropriately defined.  
\end{abstract}

\section{Introduction}

Geophysical processes often exhibit an irregular behavior\ that is due to
deterministic chaos and/or to the presence of many relevant degrees of freedom.
Striking examples are climate dynamics (Fraedrich 1988, Nicolis et al. 1997) 
and
transport processes in the atmosphere and in the oceans where
many different characteristic temporal scales are involved (Samelson 1992, 
Yang 1996). 
After Lorenz's
seminal work (Lorenz 1963), it is now well recognized that even deterministic
systems with as few as three degrees of freedom can have complex temporal
evolution  similar to genuine stochastic processes 
(Ott 1993, Beck and Schl\"ogl 1993). For
example, the large-scale and long-time dispersion characteristics of drifters
deterministically advected by a non-stochastic velocity field, giving rise to
Lagrangian chaos, can, sometimes, be well described by random walk models
(Ottino 1989, Crisanti et al. 1991). 
On the other hand, a simple probabilistic description, e.g., one in
terms of autoregressive models and Markov chains (Vautard et al. 1990, 
Nicolis et al. 1997), 
is often not able to catch the
richness of most geophysical phenomena. 

In a recent paper 
(Cencini et al. 1999),
mixing and transport in a meandering jet were investigated and compared with a
 Markov chain. It was found that, due to the presence of very
different characteristic time scales, this simple approach was not fully
satisfactory. In fact, the ubiquitous presence of a hierarchy of time scales
forces one to develop nonstandard techniques for data analysis and for their
modeling. 

The main aim of this Note is to show how a suitable Markov chain
with a small number of states can reproduce in a proper way the main
statistical features of systems similar to the one that is considered below.
The basic point is an appropriate definition of the states of the system, 
such that long-time memory effects are straightforwardly taken into 
account in a Markov chain approximation of the dynamics.  
We expect this kind of analysis to be of interest in the more general context
of systems with multiple time scales.

\section{Transport in a meandering jet}

The test case under consideration consists of a 2D meandering jet, formerly
introduced as a model for studying Lagrangian transport across the Gulf Stream
 (Bower 1991, Samelson 1992). 
In this model, the steady flow is described by the
following stream function:
\begin{equation}
\Psi(x,y)=-\tanh\left\{  k\frac{y-B\cos\left(  kx\right)  }{\sqrt{1+B^{2}k^{2}
\sin^{2}\left(  kx\right)  }}\right\}  +cy
\label{eq:steadyflow}%
\end{equation}
where $x$ and $y$ are the fluid particle zonal and meridional coordinates, 
respectively, $k$ is the
wavenumber of the flow along the $x$-direction, $B$ controls the amplitude of
the meanders and $c$ is the retrograde, counter stream, velocity component far
from the jet's core which is centered on $y=0$. The advective time, 
$\tau_{adv}$, is defined 
as the time needed for a particle, traveling close to the jet core, 
to move along the $x$ direction  a distance 
equal to the wavelength of the flow 
$\lambda = 2 \pi / k$. For the typical values of the parameters used in 
our calculations, the value of this characteristic time is   
$\tau_{adv} \simeq 4 \pi$. The recirculation time, $\tau_{rec}$, is defined 
as the period of a trajectory moving on a closed orbit next to the 
boundary of a gyre. This time is 
$\tau_{rec} \sim 10 \cdot \tau_{adv}$.   
The 2D incompressible velocity
field $(u,v)$ and the stream function $\Psi(x,y)$ are related by,
\begin{equation}
u=-{\frac{\partial\Psi}{\partial y}},\;\;\;\;v={\frac{\partial\Psi}{\partial
x}}.
\label{eq:vel}%
\end{equation}
The flow is invariant under north-south inversion ($y$ direction) 
and is periodic along
the $x$-axis with period $\lambda$. 
In the absence of perturbations, passive particles follow the
streamlines at constant $\Psi$. The steady flow corresponding to this stream
function consists of open streamlines, to be identified as ``the jet'', and
closed streamlines, to be called ``the gyres'', see Figure~
\ref{fig:flow}. Non-trivial behavior appears as soon as a perturbation is
added to the above-described flow. In particular, cross-stream transport,
i.e., from gyre to gyre, occurs either by the action of a random forcing or
due to non-stochastic, time dependent terms inserted in the stream function
which lead to the onset of Lagrangian chaos. In the stochastic model the
position of a passive particle evolves according to
\begin{equation}
\frac{dx}{dt}=u(x,y)+w_{1}(t),\;\;\;\;\frac{dy}{dt}=v(x,y)+w_{2}%
(t)\label{eq:noise}%
\end{equation}
where the stochastic fields $w_{i}(t)$ have red-noise correlations, i.e.,
\begin{equation}
<w_{i}(t)w_{j}(t^{\prime})>=2\sigma^{2}\delta_{ij}e^{-|t-t^{\prime}|/\tau
},\;i,j={1,2,}\label{eq:rednoise}%
\end{equation}
with $\sigma^{2}$ the variance of the stochastic fields and $\tau$ their
correlation time. The deterministic, chaotic model is obtained by making the
amplitude of the meanders, $B$ in equation (\ref{eq:steadyflow}), time
dependent,
\begin{equation}
B\Rightarrow B(t)=B_{0}+\epsilon\cos(\omega t+\phi)\label{eq:modB}%
\end{equation}
where $B_{0}$ is the mean meander amplitude and 
 $\epsilon$ and $\omega$ are appropriately 
chosen so as to generate large-scale
mixing, as in Cencini et al. (1999), 
i.e., leading to an overlap of resonances (Chirikov 1979); 
$\phi$ is an arbitrary
phase. 

The symbolic dynamics (Beck and Schl\"ogl 1993) 
of the noisy model can be satisfactorily approximated by
a simple minded, first order Markov process. 
In the sequel we
show that, in order to describe the chaotic dynamics via a first order Markov
chain, it is necessary to introduce  
 a more suitable definition of ``state'' which 
take into account the non trivial dynamical behavior of the system. 
This is made clear by comparing the symbolic dynamics generated by two
different partitions of the phase space, both partitions having four cells.

The simple minded partition, call it $\Pi$, is defined by the southern gyre
(symbol $1$), the southern half jet (symbol $2$), the northern half jet
(symbol $3$) and the northern gyre (symbol $4$), see
Figure~\ref{fig:partition1}. The partition takes into account the north-south
symmetry as well as the periodicity of the system. The position of the
advected particles is sampled millions of times at a time interval
$T=2\pi/\omega.$ At each sampling time, one associates one of the symbols
according to the partition cell in which the particle is located at that time.
In this way, a symbolic sequence is generated from each trajectory. Once this
has been done, one computes the transition matrix $W$, giving the
state-to-state transition probabilities in the time interval $T.$ 

While the precise nature of the perturbation, i.e.,
stochastic or purely deterministic, does not affect the gross qualitative
features of the dispersion process, it does lead to very different correlation
functions. In the deterministic, chaotic case, the transport process shows
persistent, nontrivial long term correlations which are suppressed when
stochastic noise is present. These differences are clearly seen in Figure~ 
\ref{fig:noise_chaos_sig}.
The $W_{ij}$ 
matrix element is the probability to observe state $j$ at time $(n+1)T$ 
knowing that the state $i$ occurs at time $nT$ (where $n$ is integer).   
In the
present case, using partition $\Pi,$ this analysis leads to,%

\begin{equation}
W=\left(
\begin{array}
[c]{cccc}%
0.851 & 0.147 & 0.002 & 0.000\\
0.247 & 0.535 & 0.215 & 0.003\\
0.003 & 0.215 & 0.535 & 0.247\\
0.000 & 0.002 & 0.147 & 0.851
\end{array}
\right)  .\label{eq:Wsc1}%
\end{equation}
The symmetries in the transition matrix $W$ reflect the dynamical and
statistical equivalence of the northern and southern halves of the system. For
future reference, we also quote the corresponding steady probability vector
$P=(P_{1},P_{2},P_{3},P_{4})$ which satisfies the equation  
\begin{equation}
\sum_{i=1}^4 P_i W_{ij} = P_j
\label{eq:eigenvec}
\end{equation} 
One finds: in the
gyres, \ $P_{1}=P_{4}=0.315,$ and in the jet, $P_{2}=P_{3}=0.185.$

In contraposition to the stochastic model defined by Eqs. (\ref{eq:noise}) and
(\ref{eq:rednoise}), the deterministic evolution with (\ref{eq:modB}) is
characterized by long-term correlations which are poorly reproduced by the 
Markov chain generated from $W.$
Essentially, the matrix elements $W_{23}$ and $W_{32}$ overestimate the
coupling between the two\ halves of the systems. This failure of the simple 
probabilistic model based on $W$ is
illustrated in Figure~\ref{fig:chaos_corr_smarkov} where the actual
autocorrelation functions of the states 1 and 2 are compared with the
corresponding correlation functions generated by the first order Markovian
process described by $W.$ 

As one can see, in the actual dynamics there are basically two
characteristic times, $\tau_f$ and $\tau_s$, 
that correspond to the inverse of the two exponential
decay rates at short and at long times, respectively. As order of magnitude, 
the shorter decay time is $\tau_f \sim T$, and the 
longer one is $\tau_s \sim 10^2 \cdot T$.    

A way of
estimating a lower bound for the order of the Markov chain which would be
necessary in order to adequately reproduce the system's statistical features,
see, e.g., (Khinchin 1957), is based on the block entropies $H_{n}$
which are defined by
\[
H_{n}=-\sum_{C_{n}}P(C_{n})\ln P(C_{n}),
\]
where $P(C_{n})$ is the probability of observing a sequence $C_{n}$
$=(i_{t+1},i_{t+2},.....,i_{t+n})$ of $n$ successive symbols generated by the
dynamical system, with $i_{m}\in\{1,...,4\}$ the partition cell visited at the
time $t=mT$. From the block entropies $H_{n}$ one computes next the quantities
$h_{n}=H_{n}-H_{n-1}$ which represent the average additional information
needed to specify the $n$-th symbol $i_{n}$ given the sequence $C_{n-1}$. The
limit of $h_{n}$ for $n\rightarrow\infty$ gives the Shannon entropy $h_{S}$ of
the infinite sequence $i_{1},i_{2},......$ For a Markov process of order 
$\nu$,
one has $h_{n}=h_{S}$ for all $n\geq \nu+1$, (Khinchin 1957). 
 In agreement with what was found in the paper by 
 Cencini et al. (1999), 
at least in order to reproduce the entropic properties, one would
need a Markov process of very high order. Recall that this procedure
gives only a lower bound to the order of the Markov chain.
It is clear that the strategy consisting of
simply increasing the order of the Markov process is not practical at all.
Surprisingly, we were able to give a non-obvious choice of the partition
that allows us to work with a first order Markov chain where now the time
memory effects are satisfactorily described. 
This is explained in the following. 
The new partition, denoted by $\Pi^{\ast},$ is a 4-state partition similar to
the previous $\Pi$ but the jet is partitioned in a different way: the jet
corresponds to symbol $2$ if the last previously visited cell is the southern
gyre $1$ or the cell $2$ itself; if the last previously visited cell is the
northern gyre $4$ or the cell $3$ itself, then the jet corresponds to symbol
$3$, see Figure~\ref{fig:partition2}. 
Accordingly, transitions between the two jet states $2$ and $3$ are not 
possible. 
This partition of the jet preserves
part of the long time memory of the gyre-to-current transitions, a feature of
the deterministic chaotic model which is impossible to reproduce with the
previous partition $\Pi$. In other words, this partition introduces some
memory even if formally the model remains of first order. The time signal
generated by the $\Pi^{\ast}$ Markov process is compared to the actual time
signal in Figure~\ref{fig:chaos_sig}. The memory effects are visible through
the existence of ``blocks'' describing the fast back and forth gyre-to-current
transitions in one half of the system, before moving to the other half. The
new transition matrix $W^{\ast}$ is found to be,
\begin{equation}
\;W^{\ast}=\left(
\begin{array}
[c]{cccc}%
0.850 & 0.150 & 0.000 & 0.000\\
0.243 & 0.749 & 0.000 & 0.008\\
0.008 & 0.000 & 0.749 & 0.243\\
0.000 & 0.000 & 0.150 & 0.850
\end{array}
\right)  .\label{eq:Wsc2}%
\end{equation}
The corresponding time-independent probability vector $P^{\ast}$ is equal to
the one found for the $\Pi$ partition, $P^{\ast}=P,$ as it should be.
Notice that, because the transitions between states $1$ and $2$ ($3$ and $4$) 
are much more frequent than those from state $3$ to $1$ (from $2$ to $4$), 
the matrix $W^{\ast}$ is not far from being block-diagonal. 
 
The improvements introduced by the $\Pi^{\ast}$ partition show up in the
correlation functions it generates, as can be seen in
Figure~\ref{fig:chaos_corr_hmarkov}, and in the fast convergence of the
$h_{n}$ entropies shown in Figure~\ref{fig:shannon}. 
Moreover, with the $\Pi^{\ast}$
partition the two characteristic times are well estimated.

\section{Conclusions}

We have shown that the long-term memory effects which are present in the
chaotic dispersion processes generated by the flow defined by Eqs.
(\ref{eq:steadyflow}) and (\ref{eq:modB}) can be reproduced by a 
Markov chain, provided that the partition of the phase space is performed in a
special way. 
It is important to remark that first order in a Markov process generated with 
the $\Pi^*$ partition corresponds to a very large order, 
say $\sim O(\tau_s/T)$, 
in a Markov process generated from the $\Pi$ partition. 
We hope that the ideas presented in this Note will be of use in a
more general context and we plan to apply the above discussed technique to the
analysis of other geophysical phenomena. Needless to say, when more than two
characteristic times are relevant, partitions with a larger number of cells
will be required.

\section{Acknowledgements}

R.A.P. acknowledges the hospitality of the Roman TNT group at University 
``La Sapienza''. 
A.V. acknowledges support 
from the INFM {\em Center for Statistical Mechanics and Complexity}.


\newpage

\centerline{FIGURE CAPTIONS}

\begin{figure}[ptb]
\includegraphics[angle=-90, width=1.0\textwidth]{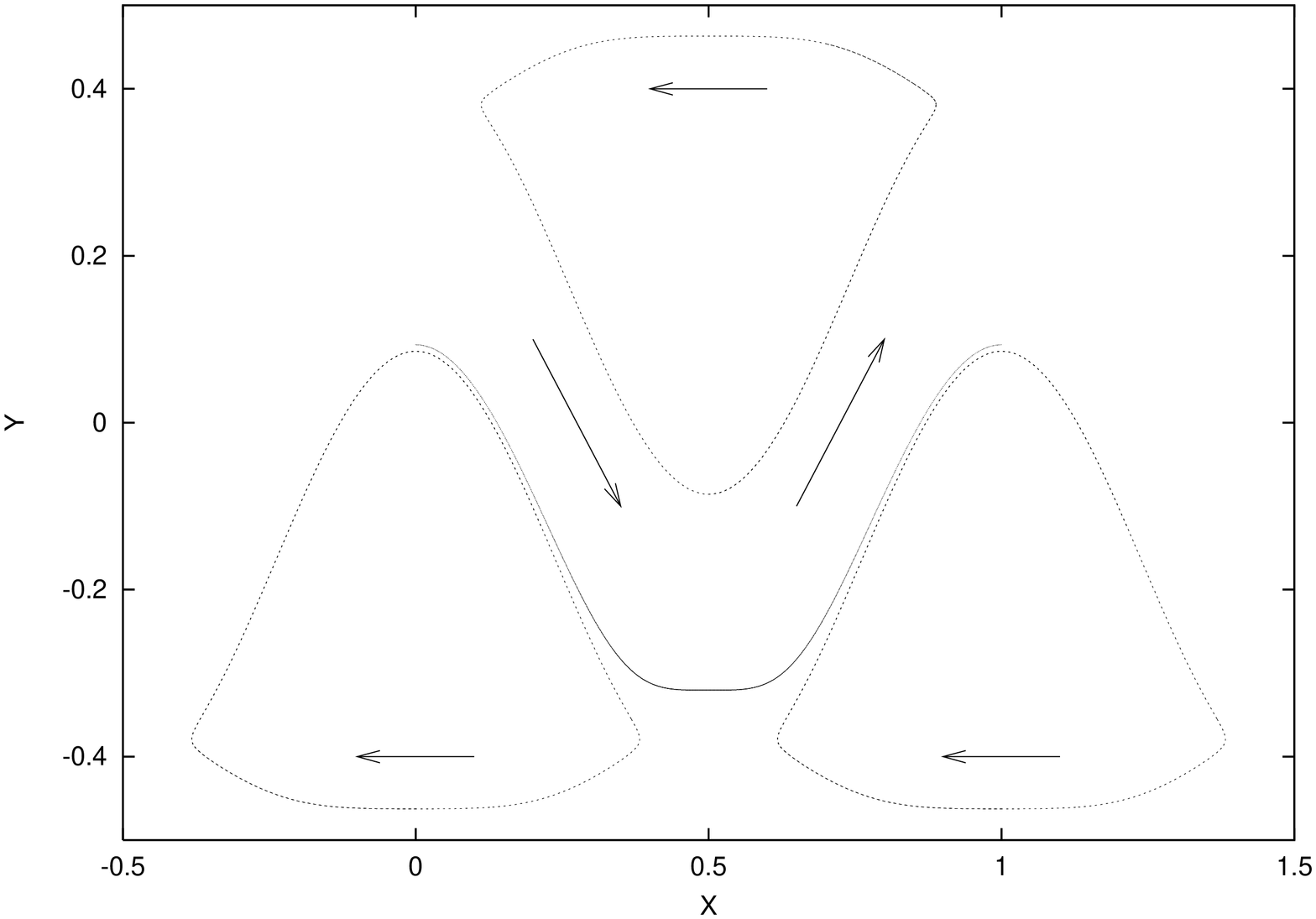}
\caption{
 }%
\label{fig:flow}%
\end{figure}

\noindent 
FIGURE~\ref{fig:flow}. 
Schematic diagram of the circulation in the meandering jet model. 
Arrows indicate the direction of particle motion: ``westerly'' advection 
in the jet current, clockwise and anti-clockwise recirculation in the 
southern and northern gyres, respectively. Spatial coordinates $X$ and $Y$ 
are given in unit of the wavelength $\lambda=2\pi/k$ and $B_0$ is 
set equal to $0.16 \lambda$, see eq. 
(\ref{eq:steadyflow}).

\begin{figure}[ptb]
\includegraphics[angle=-90, width=1.0\textwidth]{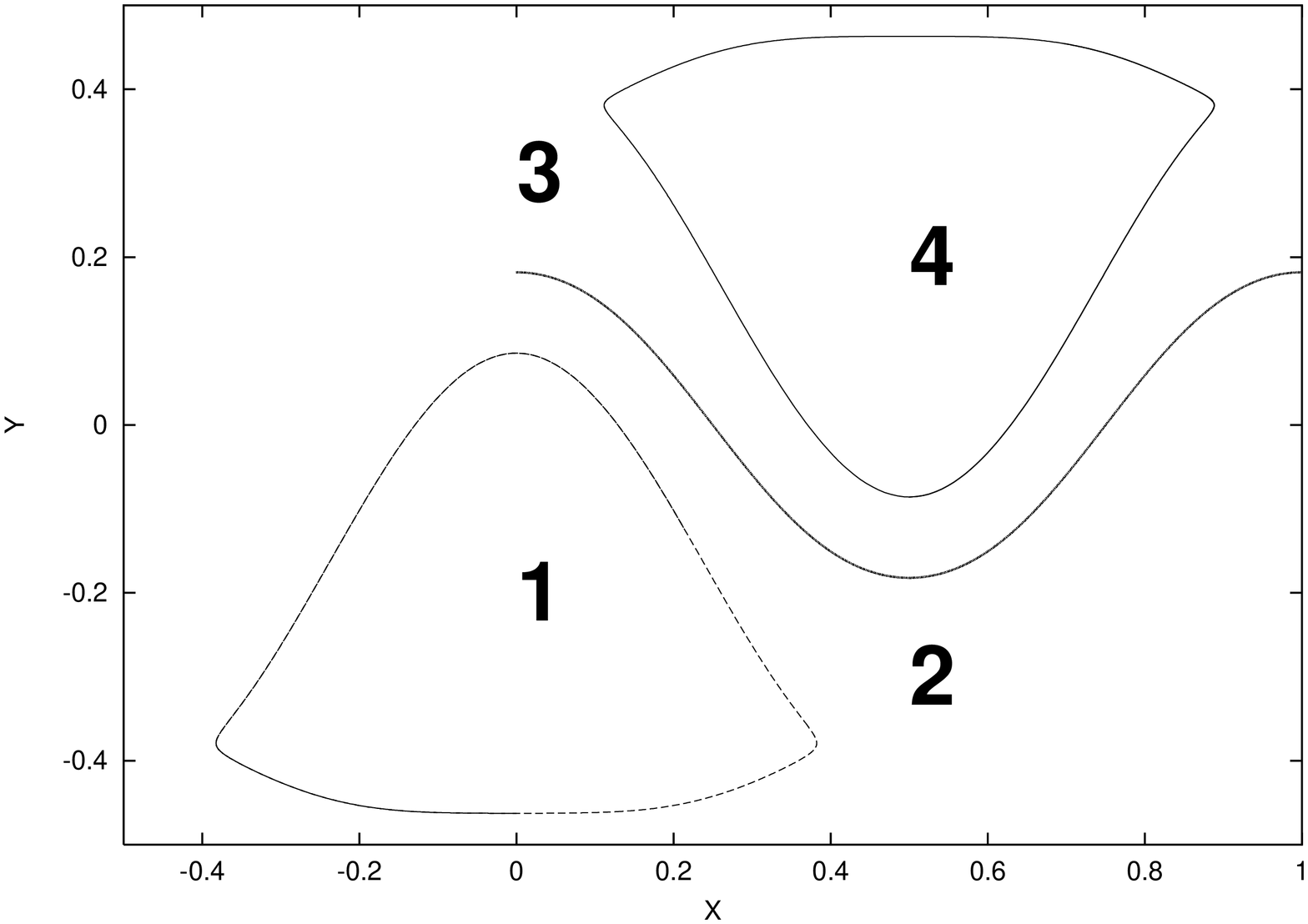}
\caption{
}
\label{fig:partition1}%
\end{figure}

\noindent 
FIGURE~\ref{fig:partition1}. 
$\Pi$
partition of the steady meandering jet: 1) south gyre; 2) south half-jet; 3)
north half-jet; 4) north gyre.

\begin{figure}[ptb]
\centerline{a)}
\includegraphics[angle=-90, width=0.75\textwidth]{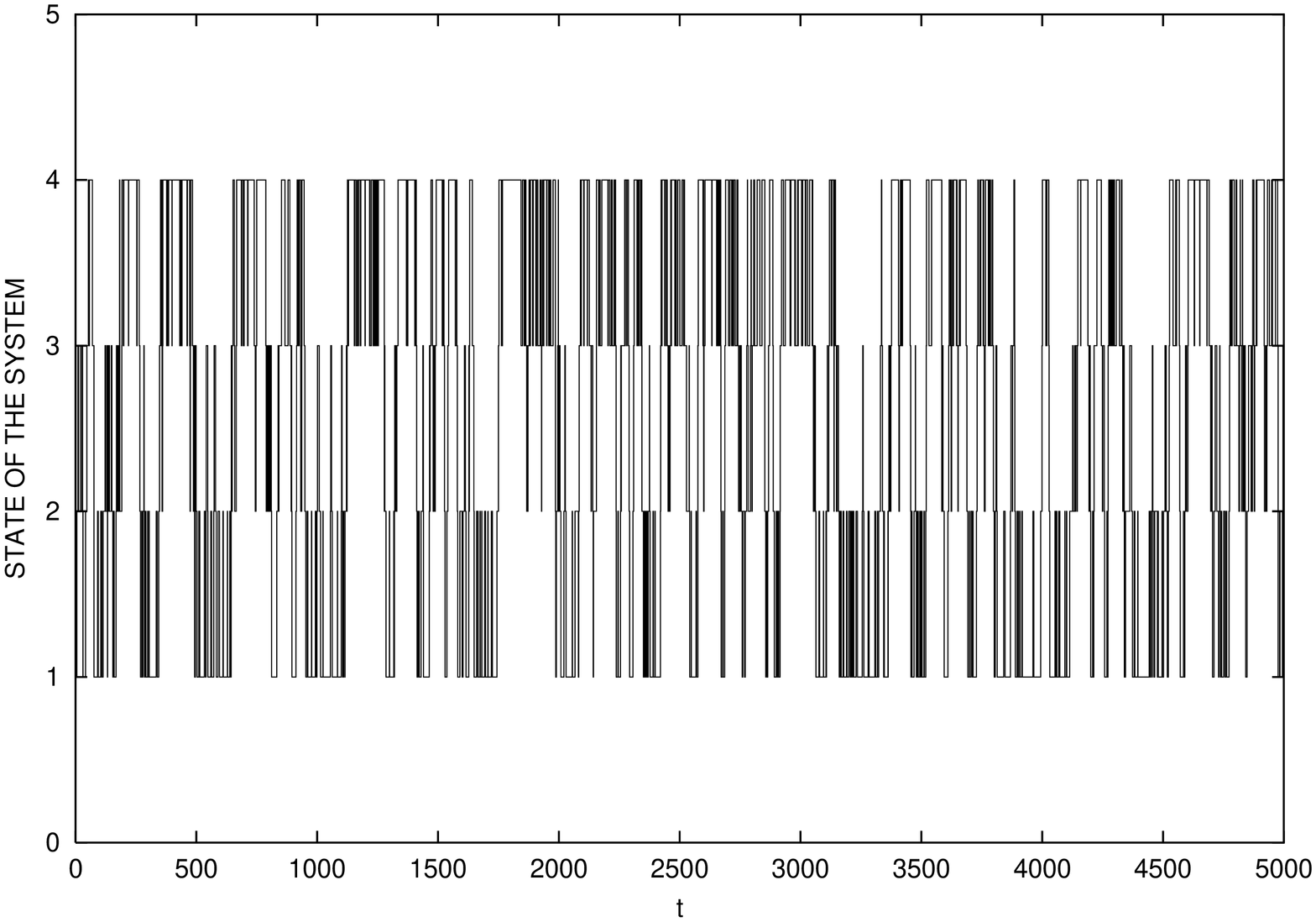}%

\centerline{b)}
\includegraphics[angle=-90, width=0.75\textwidth]{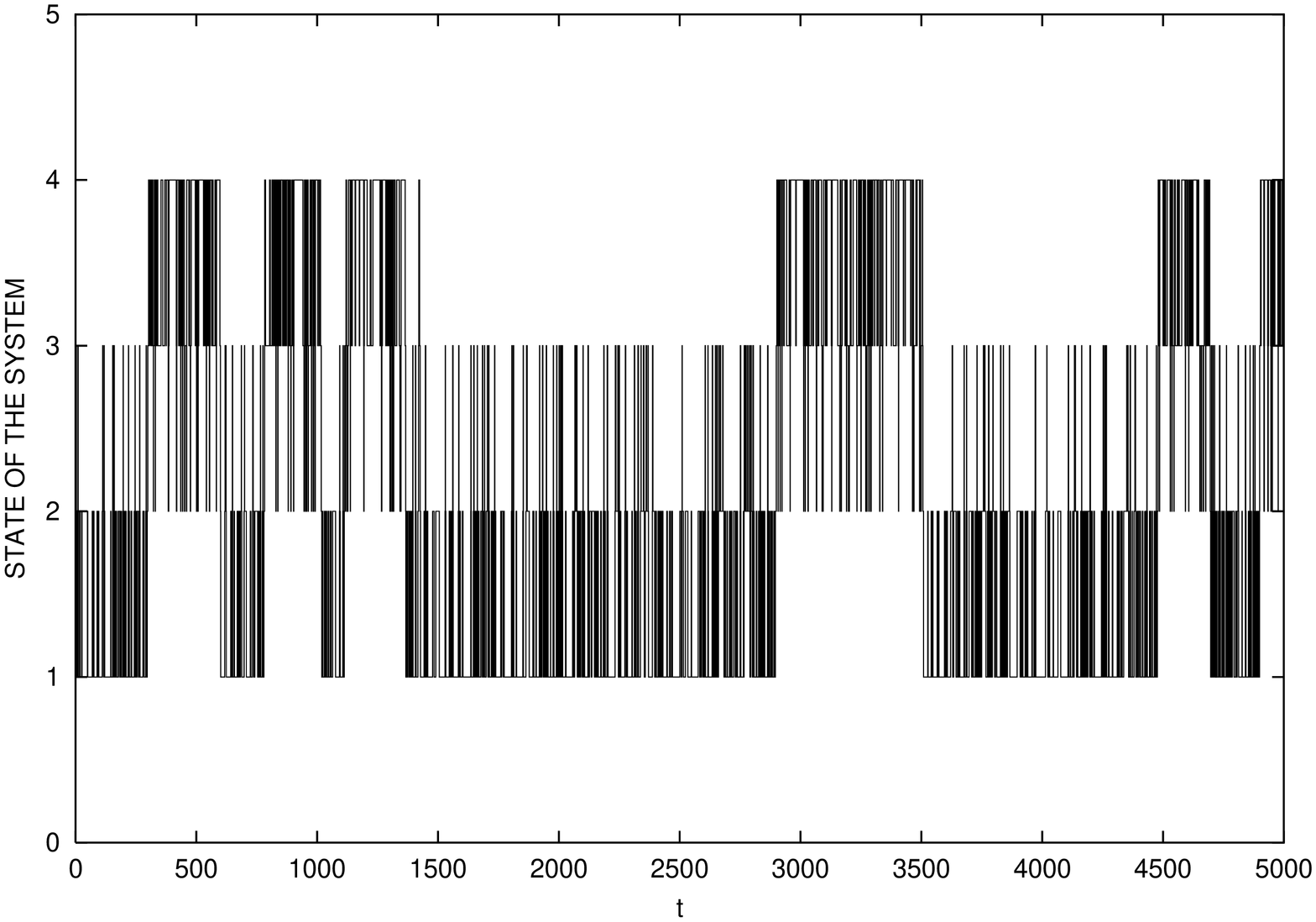}%
\caption{
}
\label{fig:noise_chaos_sig}%
\end{figure}

\noindent 
FIGURE~\ref{fig:noise_chaos_sig}. 
Comparative plot of the two typical symbolic sequences obtained from:
 a) the stochastic model and b) the chaotic model, with the parameters
 in eq. (\ref{eq:modB}) being $\epsilon = B_0 / 4$ and 
$\omega \simeq 2 \pi / \tau_{adv}$. The parameters of the red noise in 
eq. (\ref{eq:rednoise}) are $\sigma^2=6 \cdot 10^{-2} (\lambda/\tau_{adv})^2$ 
and $\tau=T/4$.

\begin{figure}[ptb]
\includegraphics[angle=-90, width=1.0\textwidth]{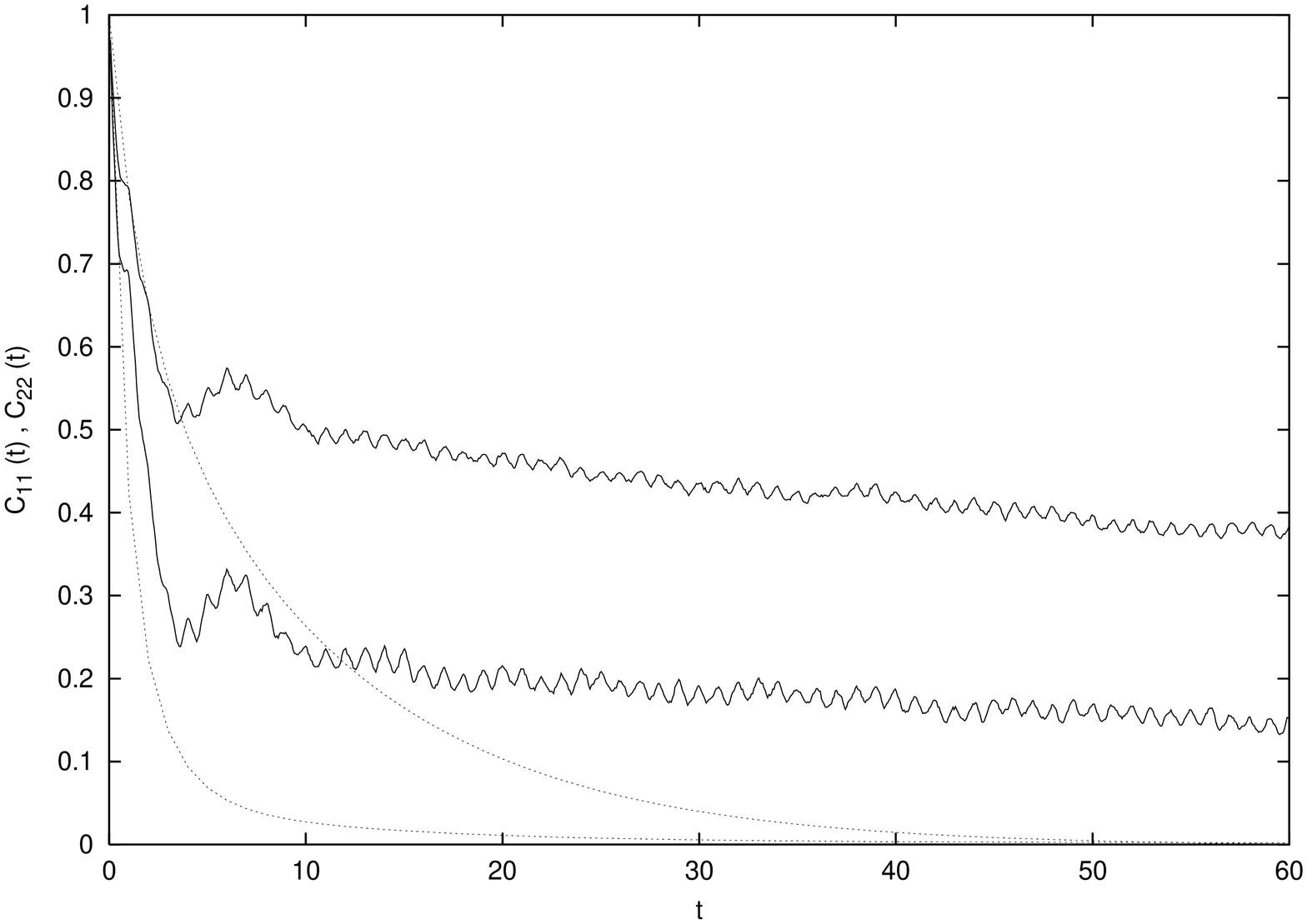}%
\caption{
}
\label{fig:chaos_corr_smarkov}%
\end{figure}

\noindent 
FIGURE~\ref{fig:chaos_corr_smarkov}. 
Autocorrelation functions for the states $1$ and $2$ of the $\Pi
$ partition, see Figure~\ref{fig:partition1}: actual autocorrelations
from the chaotic model (full lines) 
and those generated by the transition-rate matrix 
$W$ (dotted lines). Parameters of
the deterministic perturbation: as in Figure~\ref{fig:noise_chaos_sig}. 

\begin{figure}[ptb]
\includegraphics[angle=-90, width=1.0\textwidth]{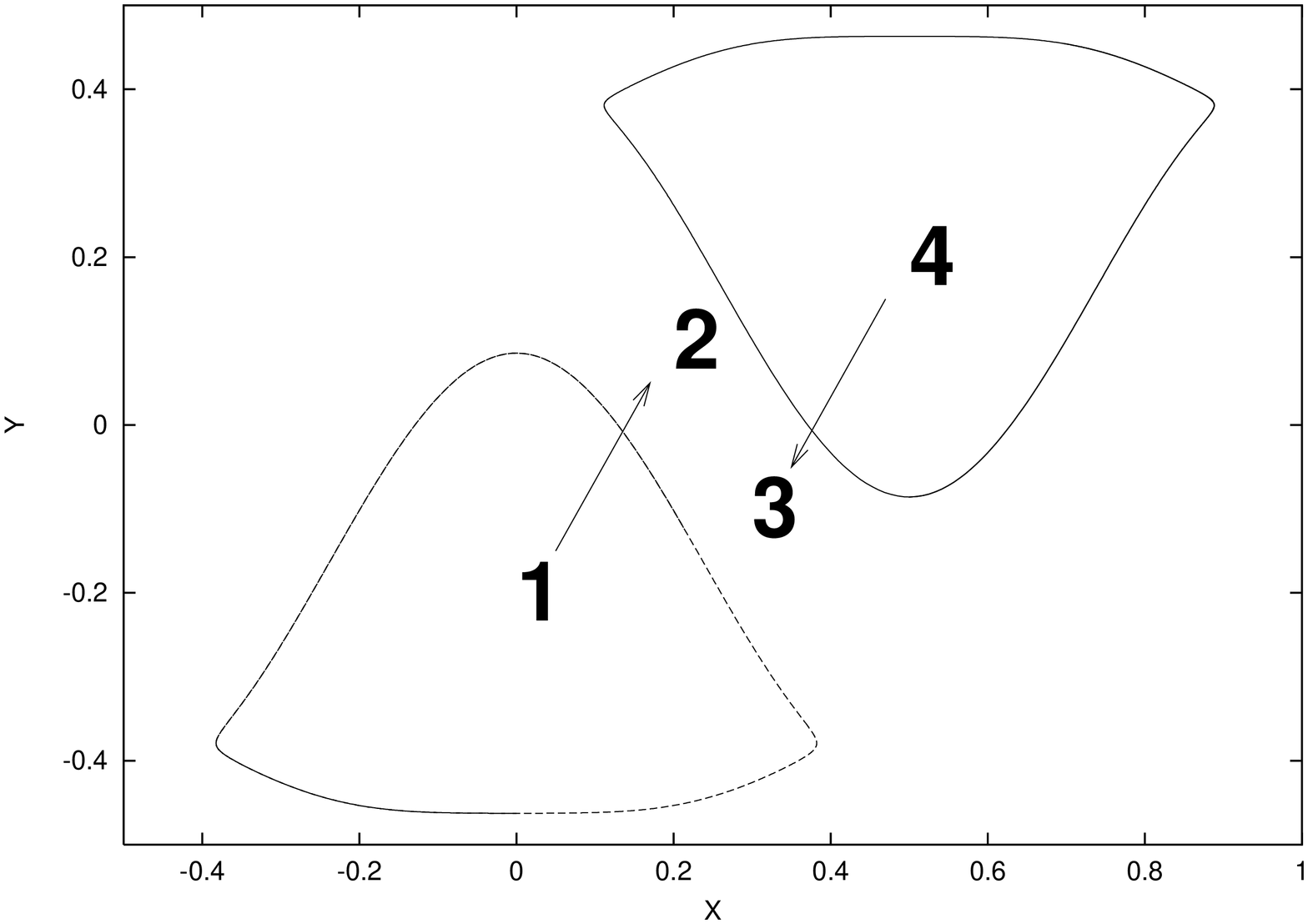}
\caption{
}
\label{fig:partition2}%
\end{figure}

\noindent 
FIGURE~\ref{fig:partition2}. 
$\Pi
^{\ast}$ partition of the steady meandering jet: 1) south gyre; 2) and 3) jet;
4) north gyre. The central jet corresponds to state $2$ if the last visited
gyre is state $1$, and to state $3$ if the last visited gyre is state $4$.

\begin{figure}[ptb]
\centerline{a)}
\includegraphics[angle=-90, width=0.75\textwidth]{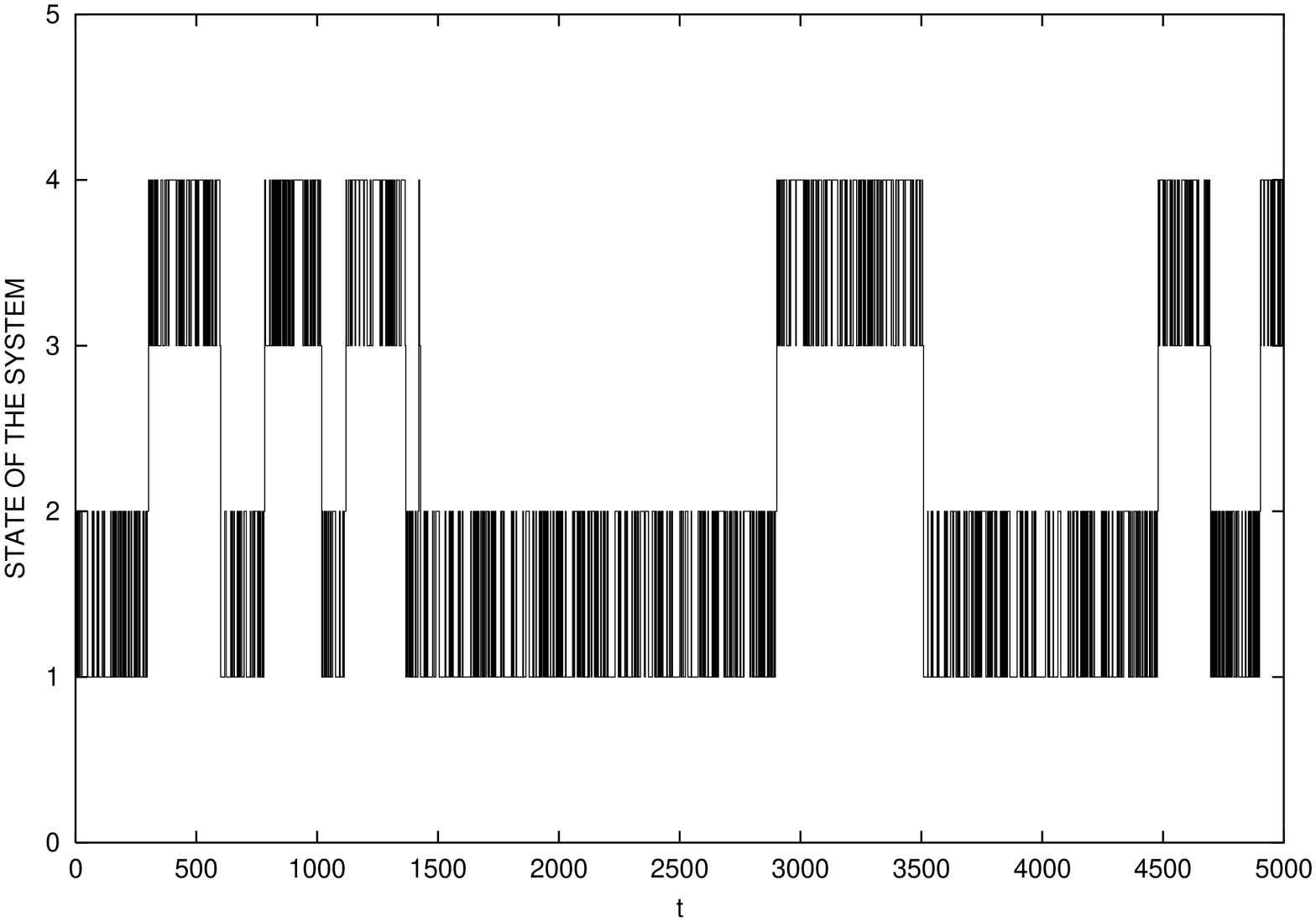}%

\centerline{b)}
\includegraphics[angle=-90, width=0.75\textwidth]{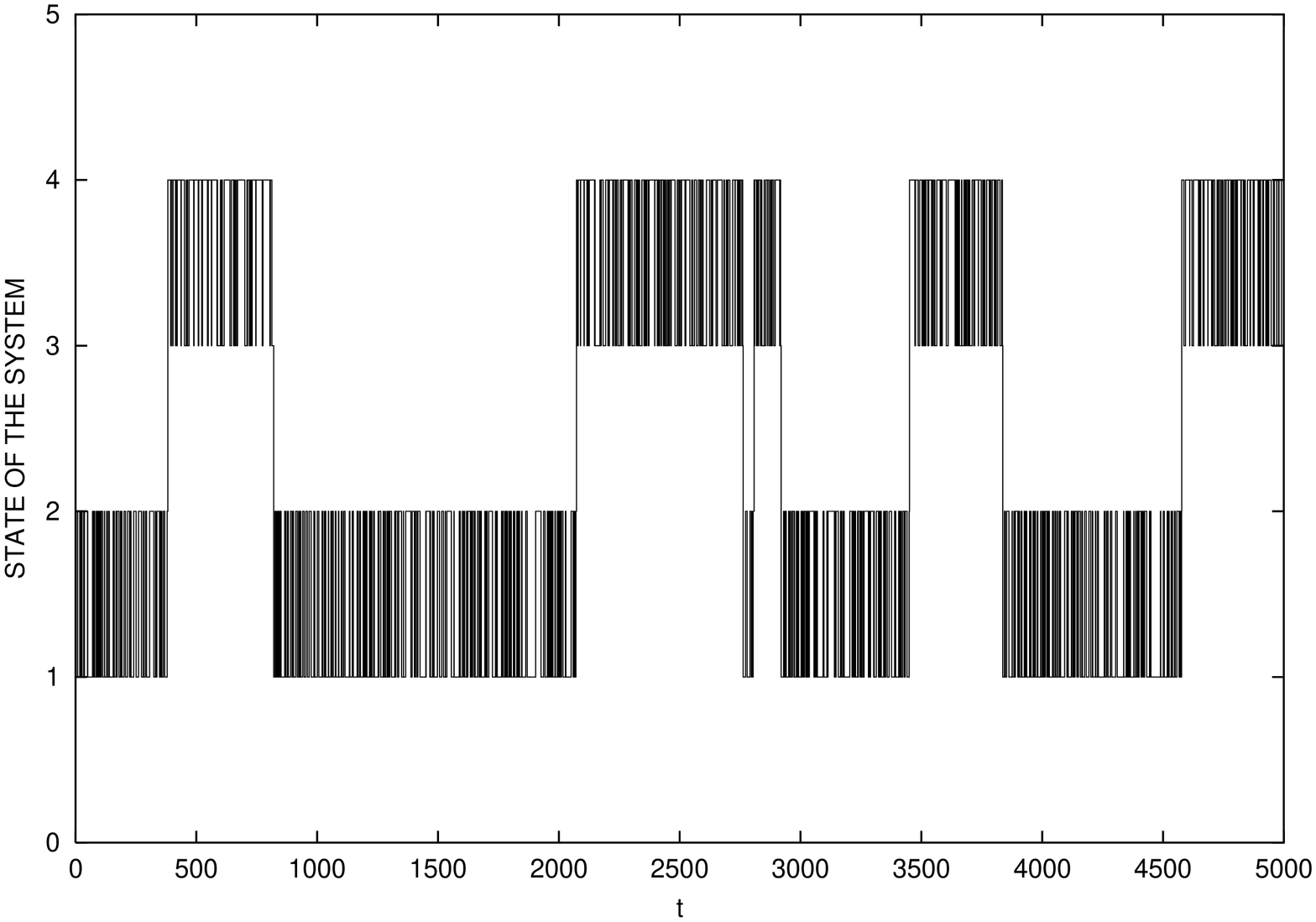}%
\caption{
}
\label{fig:chaos_sig}%
\end{figure}

\noindent 
FIGURE~\ref{fig:chaos_sig}
Comparative plot of: a) the chaotic model symbolic dynamics using 
the partition  
$\Pi^*$ and b) the symbol sequence simulated on the basis of the 
transition-rate matrix 
$W^{*}$.  

\begin{figure}[ptb]
\includegraphics[angle=-90, width=1.0\textwidth]{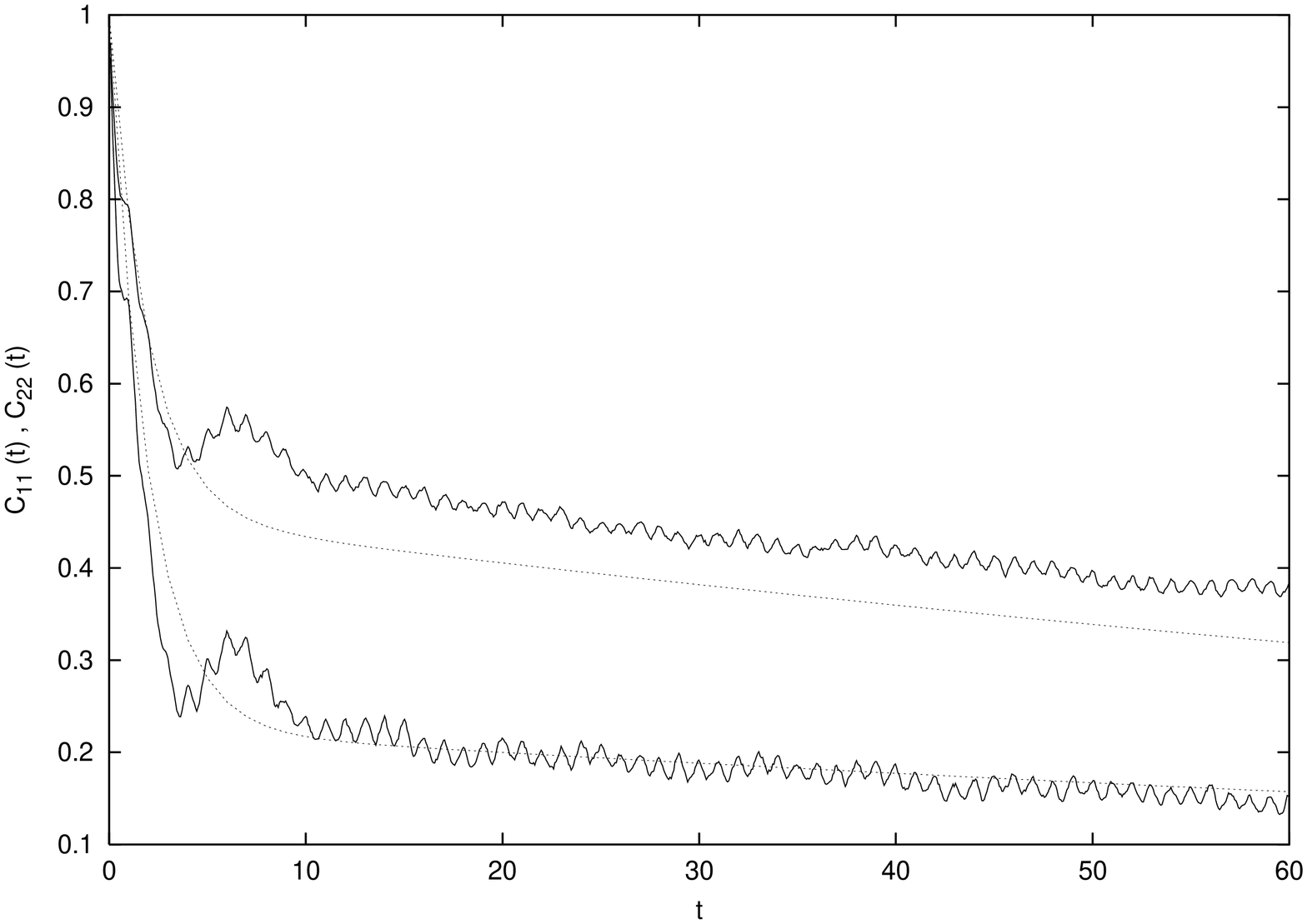}%
\caption{
}
\label{fig:chaos_corr_hmarkov}%
\end{figure}

\noindent 
FIGURE~\ref{fig:chaos_corr_hmarkov}. 
Autocorrelation functions for the states $1$ and $2$ of the 
$\Pi^{*}$
 partition, see Figure~\ref{fig:partition2}: actual autocorrelations from
the chaotic model (full lines) and from the Markovian approximation 
 generated with $W^{*}$ (dotted lines).  

\begin{figure}[ptb]
\includegraphics[angle=-90, width=1.0\textwidth]{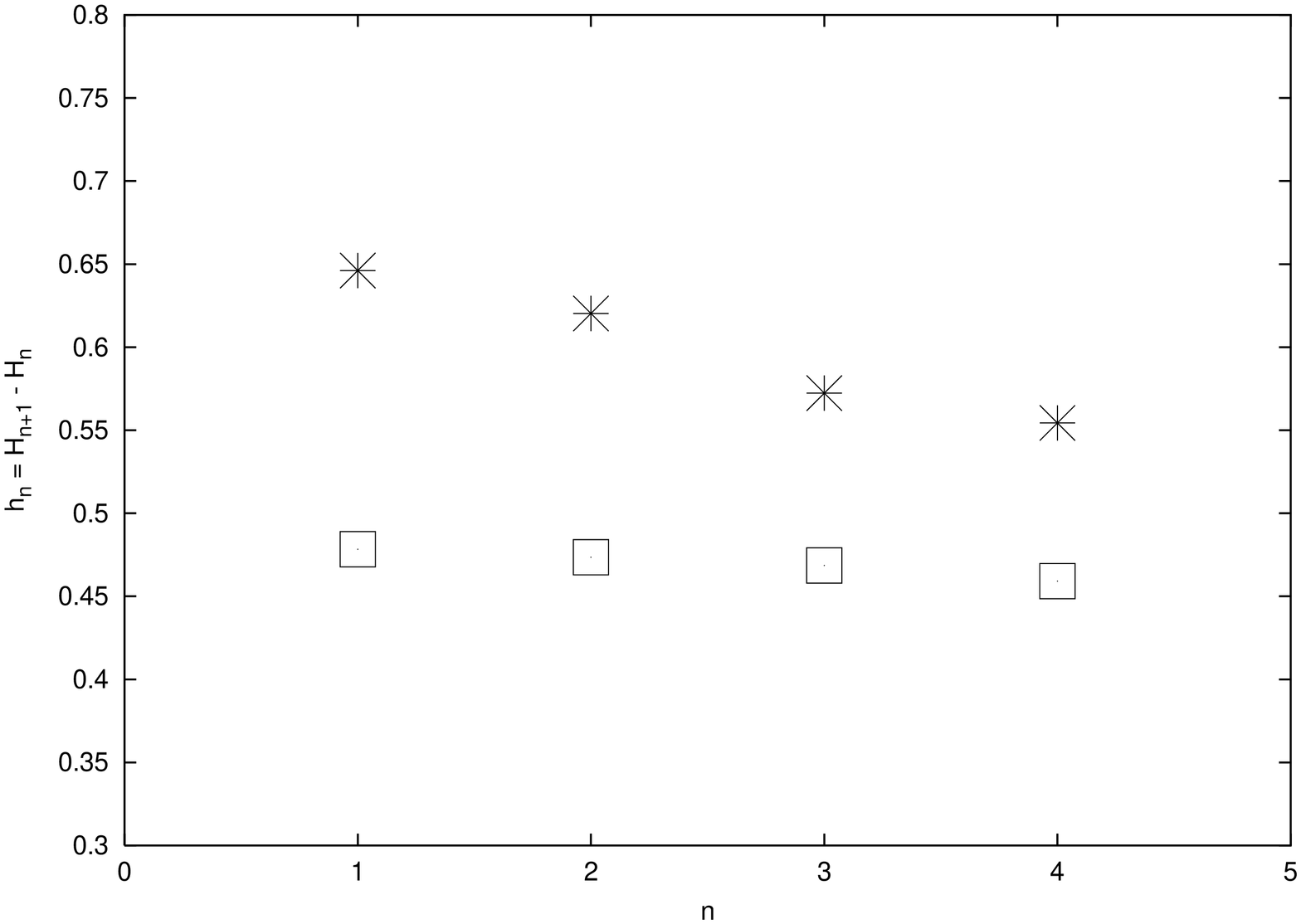}%
\caption{
}
\label{fig:shannon}%
\end{figure}

\noindent 
FIGURE~\ref{fig:shannon}. 
Differential Block Entropy $h_n = H_{n+1}-H_{n}$, as a function 
of the sequence-length $n$, for 
the $\Pi$ ($\star$) and the $\Pi^*$ ($\square$) partitions. 
For each length $n$ approximately 
 $\sim 10^7$ sequences were used.
Notice the very different rate of convergence of $h_n$ to its limit 
value (Shannon entropy) $h_S$ in the two cases.

\newpage

\pagebreak[4]

\pagebreak[4]

\pagebreak[4]

\pagebreak[4]

\pagebreak[4]

\pagebreak[4]

\pagebreak[4]

\end{document}